# Touchscreen Voting Machines Cause Long Lines and Disenfranchise Voters

William A. Edelstein[1] and Arthur D. Edelstein[2]
*Draft version October 30, 2008*

*Computerized touchscreen "Direct Recording Electronic" DRE voting systems have been used by over 1/3 of American voters in recent elections. In many places, insufficient DRE numbers in combination with lengthy ballots and high voter traffic have caused long lines and disenfranchised voters who left without voting. We have applied computer queuing simulation to the voting process and conclude that far more DREs, at great expense, would be needed to keep waiting times low. Alternatively, paper ballot-optical scan systems can be easily and economically scaled to prevent long lines and meet unexpected contingencies.*

The controversial Presidential election in 2000 convinced Congress that US voting technology should be upgraded, and the result was the Help America Vote Act (HAVA) passed in 2002 [1]. This legislation established various rules for voting systems, included provisions to make voting accessible to people with a wide range of disabilities, and funded states to buy new voting equipment.[1][2]

Most states and voting precincts now have either computer touchscreen "Direct Recording Electronic" (DRE) systems (38% of voters in 2006) or paper ballot-optical scan (PBOS) equipment (49% of voters) [2].

DREs generally use a touch-screen on which voters enter their choices electronically (e.g. [3]). Votes are recorded digitally on a memory card and tabulated at the end of the voting day.

With PBOS systems, voters use a pen or pencil to fill in circular or elliptical "bubbles" or complete a line on a paper ballot (e.g. [4]). Completed ballots are fed through a scanner which tallies the votes. The voter-marked ballots are subsequently available for manual or machine recounts or audits.

DREs record votes and keep track of vote totals. Unfortunately, it is not possible to recount or audit paperless DREs and votes have been lost or questioned because of DRE malfunctions [5]. As a result of DRE breakdown and security concerns, there has been a growing demand for "paper trails" for DREs to enable recounts and audits [6, 7].

Another serious problem associated with DREs is the formation of long lines of voters in many venues across the United States (California, Florida, Maryland, Mississippi, Ohio, Pennsylvania, Tennessee, Utah and elsewhere [6, 8-14]), sometimes requiring several-hour-long waits to vote. Inevitably, some voters caught in such situations—for example, the elderly, people with disabilities or illness, people needing to get back to work, parents needing to care for children—leave without voting and are thereby disenfranchised [14]. The simple reason that these delays occur is that there are not enough DREs at each precinct to allow voting in a timely manner.

In contrast, PBOS systems can easily be expanded to meet an unexpectedly large number of voters or to allow extra time to mark a complex or long ballot. For a PBOS system, the equivalent traffic choke points to DREs are inexpensive marking stations that may be as simple as a cardboard screen taped to a table. Additional privacy screens can be immediately installed if a need for them becomes apparent. In other words, PBOS systems allow for "just-in-time" ballot stations not possible with DRE systems.

It is intuitively evident that there must be an ample capacity of voting stations in order to cope with unexpected fluctuations in voter numbers or voting time. Thus it is important to understand the interaction between voting systems and voting patterns.

We have used queuing simulation of elections to study voter flow as a function of voter numbers and time to vote [15, 16]. We have derived a "Queue-Stop" rule that that can avoid the formation of significant lines. This is easy to accomplish with PBOS, but prohibitively expensive for DREs.

As an example, we apply this approach to Maryland, which presently uses Diebold Accuvote OS touchscreen DRE voting machines. Maryland has nearly 1,800 voting precincts containing from 19 to 7,000 registered voters each with an average of 1,740 voters [17]. Maryland state regulations require one DRE for each 200 registered voters, plus an additional voting unit for every fractional part of that number." [18] The number of DREs per precinct ranges from 2 to 35 with an average of $9.2 \pm 4$, and approximately 16,500 of these DREs are used in every election. [17].

---

1. Department of Radiology, Johns Hopkins University, Baltimore, MD, 21218, w.edelstein@gmail.com
2. Department of Physics, University of California, Berkeley, CA 94720, arthuredelstein@gmail.com

We consider a Maryland election in which individual voting takes an average of 5 minutes and there is a 75% turnout, i.e. 150 voters per DRE. Maryland has a 13-hour Election Day starting at 7 a.m. and ending at 8 p.m. We assume three heavy traffic periods—7-9 am, 12-2 pm, and 5 to 8 pm—and suppose that 10% of voters come in each hour during these intervals, while 5% per hour arrive during the rest of the day. We derive wait time statistics by simulating 10,000 elections, assuming a Poisson voter arrival process for the average rates described above. These voter traffic variations are consistent with observations in Columbia County, NY [19, 20].

Figure 1 shows queuing simulation results for four Election Days with maximum waiting times or late closing times over 50 minutes. The long delays occur during heavy voter traffic periods: morning, lunch and evening.

One might ask whether the maximum wait times or closing delay could be a fluctuation for only a few voters, but this is not the case. It is evident that buildup and decay of waiting times—the development and contraction of extensive lines—is slow, so that a high maximum wait implies a drawn-out election experience for many voters. For example, the four plots in Fig. 1 have 10%-20% (150-300) of all voters waiting over 30 minutes.

Voting congestion is analogous to highway traffic jams. When car numbers are low, traffic flows freely. As vehicle numbers increase, traffic slows gradually until a density is reached at which a few cars become stationary, traffic locks up, and long lines form that can take hours to clear.

Figure 2A shows distributions of maximum waiting times (the longest time a voter waits in each of 10,000 elections) for precincts with different numbers of DREs, and Figure 2B shows distributions of late closings. The variations are a result of voter number fluctuations, and it is apparent that precincts with more DREs smooth out the variations.

We can find the fraction of precincts with specific waiting times or late closing delays by determining the fractional area under each curve in Fig. 2 starting with the time of interest. For example, 82.5% of precincts with 2 DREs will have maximum waits of more than 45 minutes compared to 59.1% of 10-DRE precincts. 63.2% of 2-DRE precincts will have greater than 45-minute overtimes compared to 68.6% of 10 DRE precincts. Tables 1A and 1B show these values

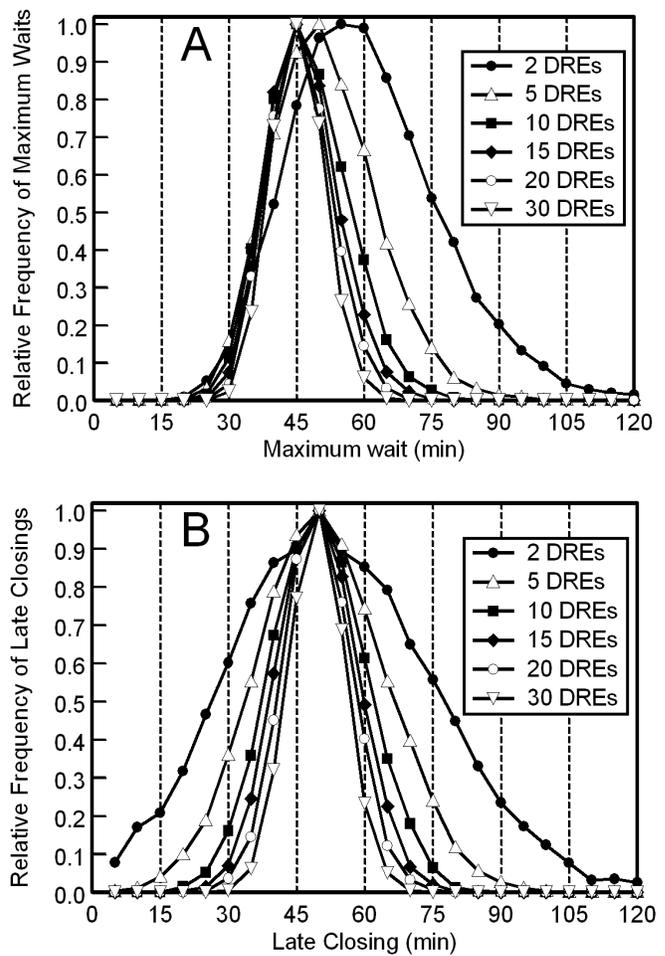

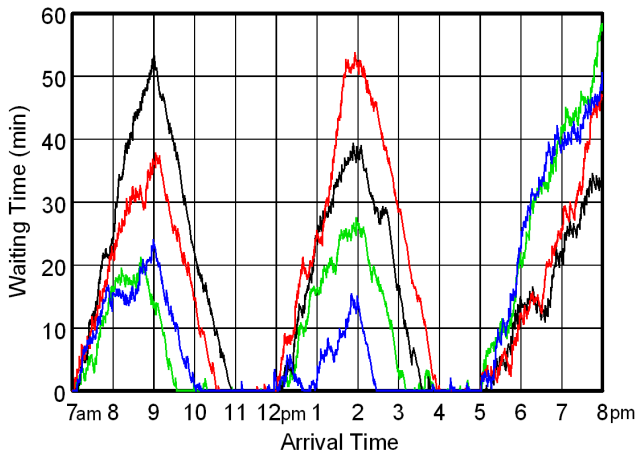

**Figure 1**. Four election sessions with maximum waiting times over 50 minutes. These occur during morning, lunch or evening heavy voter flow periods. Note that the buildup and decay of long waits—in other words, long queues—is slow, so a long maximum wait is an indication that many voters will have long delays.

**Figure 2**. (A) Maximum wait and (B) late closing times for a precinct with 150 actual voters per machine in a 13 hour Election Day. 10% of the voters arrive each hour between 7-9 am, 11 am-1 pm, and 5 pm – 8 pm. 5% of the voters arrive during each of the other six hours. 10,000 elections were simulated and the results normalized so that the maximum point has value 1. More machines smooth fluctuations and produce narrower distributions, even though there are still 150 voters per machine.

**Table 1A.** Fraction of precincts that will have the maximum waiting times specified as a function of the number of DREs in the precinct. Statistics were calculated from 10,000 simulated elections assuming 150 voters per DRE, each taking an average of 5 minutes to vote, with a 13 hour Election Day.

|         | >15 min | >30 min | > 45 min | > 60 min | > 75 min | > 90 min | > 105 min | > 120 min |
|---------|---------|---------|----------|----------|----------|----------|-----------|-----------|
| 2 DREs  | 100.0%  | 98.6%   | 82.5%    | 47.4%    | 18.3%    | 5.6%     | 1.2%      | 0.3%      |
| 5 DREs  | 100.0%  | 98.6%   | 69.2%    | 21.3%    | 2.9%     | 0.2%     | 0%        | 0%        |
| 10 DREs | 100.0%  | 99.0%   | 59.1%    | 9.6%     | 0.3%     | 0%       | 0%        | 0%        |
| 15 DREs | 100.0%  | 99.4%   | 54.8%    | 5.1%     | 0%       | 0%       | 0%        | 0%        |
| 20 DREs | 100.0%  | 99.7%   | 53.6%    | 2.6%     | 0%       | 0%       | 0%        | 0%        |
| 30 DREs | 100.0%  | 99.9%   | 51.5%    | 0.7%     | 0%       | 0%       | 0%        | 0%        |

**Table 1B.** Fraction of precincts that will have long closing delays specified as a function of the number of DREs in the precinct. Statistics were calculated from 10,000 simulated elections on Election Day.

|         | >15 min | >30 min | > 45 min | > 60 min | > 75 min | > 90 min | > 105 min | > 120 min |
|---------|---------|---------|----------|----------|----------|----------|-----------|-----------|
| 2 DREs  | 96.6%   | 85.8%   | 63.2%    | 37.1%    | 16.7%    | 5.7%     | 1.4%      | 0.40%     |
| 5 DREs  | 99.6%   | 92.9%   | 65.3%    | 25.7%    | 4.7%     | 0.4%     | 0.02%     | 0%        |
| 10 DREs | 100.0%  | 97.6%   | 68.6%    | 17.6%    | 0.9%     | 0.02%    | 0%        | 0%        |
| 15 DREs | 100.0%  | 99.2%   | 71.6%    | 12.5%    | 0.3%     | 0%       | 0%        | 0%        |
| 20 DREs | 100.0%  | 99.6%   | 75.0%    | 9.0%     | 0.03%    | 0%       | 0%        | 0%        |
| 30 DREs | 100.0%  | 100.0%  | 79.1%    | 4.9%     | 0%       | 0%       | 0%        | 0%        |

for a series of maximum waits and closing delays.

To test the sensitivity of queue formation to changing parameters, we carried out 100,000-voter election simulations for a 10-DRE precinct varying time to vote and number of voters per DRE.

Fig. 3(A) shows the fraction of precincts with various waiting times as a function of the time needed to vote assuming (as above) precincts with 150 actual voters per DRE. Figure 3(B) displays the same fraction vs. number of voters per DRE in precincts assuming a voting time of 5 minutes.

Both these plots illustrate the extreme sensitivity of the generation of long lines/waits to polling place conditions. From Fig. 3(A), a 4.6 minute voting time would result in only 0.1% of precincts with a maximum wait of over one hour. But a 5 minute voting time would cause 10% of precincts to have one-hour waits. 138 voters per DRE in Fig. 3(B) cause 0.1% of precincts to have greater than one hour maximum waits, but 10% of precincts would have those kinds of waits with 150 voters per DRE.

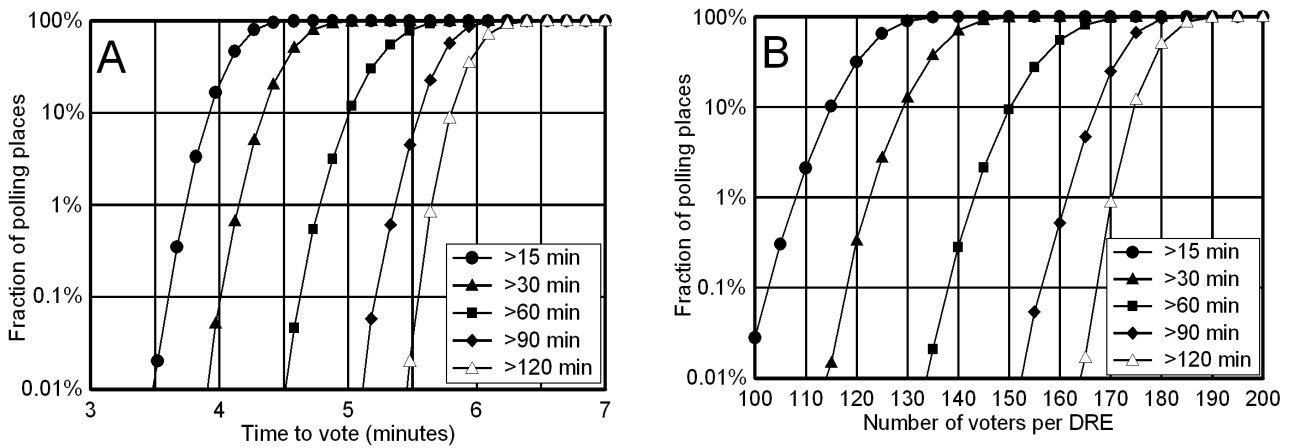

**Figure 3.** (A) Fraction of polling places with maximum waiting times vs. time to vote, given 150 voters per DRE and (B) number of voters per DRE given a 5 minute voting time. 100,000 elections were simulated for each data point. The results show that small changes in time to vote (A) or voters per DRE (B) produce big changes in the fraction of polling places with long waits.

So a 9% change of time to vote or number of voters per DRE causes a 100X increase in the number of precincts with greater than 60 minute maximum waits.

Given the sensitivity of waiting times to small changes in voter numbers and voting times, can we specify a number of DREs that will prevent queues? In general we know that such a rule must provide a substantial reserve of DREs in order to cope with highly variable election conditions.

We suggest for this purpose a "Queue Stop" rule which is calculated using the formula

$$NV_{DRE} \times T_{Vote} \leq \frac{T_{Day}}{2} \qquad (1)$$

where $NV_{DRE}$ is the number of voters per DRE in a day, $T_{Vote}$ is the number of minutes it takes each voter to mark a ballot and $T_{Day}$ is the total minutes in the Election Day.

For example, suppose we have a 13-hour voting day (780 minutes) and voting takes on average 5 minutes. This gives $NV_{DRE} \leq (1/2) \times (780/5) = 78$. So there should be at least one DRE for every 78 actual voters.

Assuming a potential 75% turnout, the number of registered voters per DRE, $NV_{reg}$, is related to the number of actual voters per DRE by

$$NV_{DRE} = 75\% \cdot NV_{reg} \qquad (2)$$

Continuing our example, we should therefore have at least one DRE for $NV_{reg} = (78/75\%) = 104$ registered voters. This is nearly twice as many DREs as are prescribed by Maryland law, which specifies one DRE per 200 registered voters [18].

We can rearrange Eq. 1 to find a recommended average voting time for a given set of election parameters.

$$T_{Vote} \leq \frac{1}{2} \times \frac{T_{Day}}{NV_{DRE}} \qquad (3)$$

200 registered voters specified by Maryland law [18] would give $NV_{DRE} = 150$ voters for a 75% turnout. Eq. 3 says $T_{Vote} \leq (780\,\text{min}/150) \times (1/2) = 2.6\,\text{min}$. If $T_{Vote}$ exceeds this value, then long lines will likely form somewhere.

Figure 4 is a contour plot of waiting times vs. voting time and voter numbers. The closeness of the contours again indicates the sensitivity of waiting times to voter numbers and average voting times. The lowest trace is a "Queue-Stop" contour, following Eq. 1 above.

The Queue-Stop contour lies well below other curves and should therefore eliminate the chance of long queues if the combination of average voting time and number of voters per DRE are on or below that line.

However, an unexpected fluctuation—a long ballot or extra voters—can easily push the queuing product $NV_{DRE} \times T_{Vote}$ higher in the plot where long waits become probable.

The paper ballot marking station in a PBOS system represents the same potential choke point for voters as does a DRE. The high cost of DREs, however—about $3,000 each in MD [21]) —compared to inexpensive ballot marking privacy booths ($200 [22]) or cardboard screens (a few dollars) means that it is far more economical to provide a large reserve capacity for ballot marking than to do the same for DREs. More crucial is the fact that extra ballot marking capacity can be installed essentially instantly with paper ballot voting—for example, by taping extra cardboard screens to tables—whereas it is difficult to bring in extra DREs, assuming that the local election jurisdiction even has any extras.

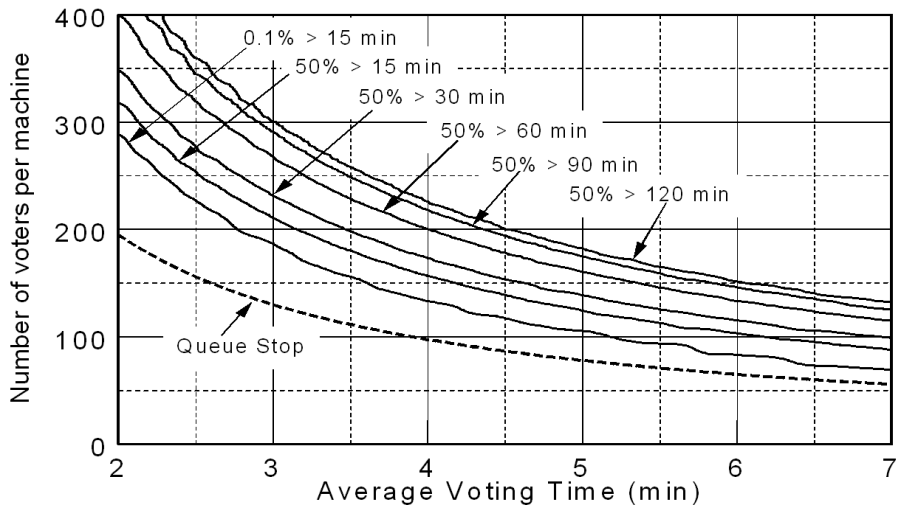

Figure 4. Maximum waits as a function of average voting time and number of voters per machine for a precinct with 10 DREs. The "Queue Stop" rule that would eliminate waits is calculated from the formula $NV_{DRE} \times T_{Vote} \leq T_{Day}/2$. Its curve lies well below the contours for even a 15-minute wait. The extreme sensitivity of maximum waiting times is again exhibited by the closeness of these contours.

Lee, Massachusetts, with 3800 active voters changed from eight mechanical lever voting machines to PBOS with 35 marking booths and one scanner. In the 2004 general election, 3200 people voted in Lee. The town clerk Suzanne Scarpa said that the lever machines in the past had caused "long, long lines," but that there were no lines for the marking booths or the scanner [23].

We can also apply queuing simulation to ballot scanning. In the voting documentary "Bought and Sold," ballots pass through two different ballot scanners in less than 1 s each [23]. The total cycle time between corresponding positions for consecutive voters must include the time to walk to and leave the scanner.

The cycle time for a very simple scanner that just accepts and processes the paper could be 5 s or less. If the voter has to look at a scanner display which indicates over- or undervotes, the time may increase, say to 10 s or more. (An "undervote" means that the voter has not made a choice in one of the ballot contests; an "overvote" occurs when a voter has improperly chosen too many candidates.) Inspecting a ballot image could take 30 s to 60 s or longer.

We have calculated the probability of maximum waiting times for various numbers of voters taking 5 s, 10 s, 30 s or 60 s to scan their ballots. Since these processes are relatively stable, we set a threshold as the number of voters per scanner that has a probability of 0.1% that there will be a maximum wait of more than 15 minutes. Results are shown in Table 2.

A single scanner with a vote cycle time of 5 s could support over 7300 voters. If two sheets of paper are needed, then the cycle time might move toward 10 s, in which case a single scanner would support about 3600 voters. (Voting scanners generally can scan both sides of a single sheet simultaneously.)

In Maryland, the largest single precinct has 6971 registered voters[17]. A 75% turnout for this precinct would be 5228 voters, well within the limits for a single, simple scanner taking 5 s per voter. A 75% turnout giving 3500 voters (10 s per voter) corresponds to 4667 registered voters. In Maryland, only 4 out of 1795 polling places have more than 4667 voters. Thus the overwhelming majority of Maryland polling places could function well with a single scanner and voter cycle time 5-10 s.

We can also consider possible queues at the check-in database terminals known as "E-Pollbooks" in Maryland. WAE has served as an election worker in Maryland for three elections and measured the average check-in time to be approximately 1 minute. Applying the last row of Table 2, we conclude that there should be at least one check-in terminal for every 462 actual voters or 616 registered voters based on 75% turnout.

Maryland has nearly 1800 polling places. If 180 polling places (10%) or 18 polling places (1%) or even 2 (0.1%) were seriously congested with long delays for voters, there could be significant effects on local, regional or national elections and consequent political disputes. As noted by Clive Thompson, "voting requires a level of precision we demand from virtually no other technology." [5]

The 2004 and 2006 Maryland elections had a number of voting precincts with very long lines. The 2006 ballot in Prince George's County had 37 items including election contests and ballot questions (aka "propositions" or "referendums"). Ms. Rebecca Wilson, a Chief Election Judge there, estimates that voting in her precinct took 15-25 minutes on average in that election. [24]

The 2008 Presidential election is hotly contested and turnout of over 80% is predicted in Maryland [25]. Some ballots will be lengthy. In addition to the Presidential, Congressional and other electoral contests, there will be two statewide ballot questions and many local ballot questions: 7 for Prince George's County, 11 for Baltimore County and 16 for Baltimore City. [26]

In order to avoid long lines it is necessary to have a large reserve capacity to deal with election fluctuations. The Maryland formula for DRE numbers *[18]* might seem reasonable. However, our calculations show that a 75% turnout, and a 5-minute or longer voting time average, would require twice as many DREs to guarantee a smooth election.

Thus conditions will be ripe Nov. 4 for long lines in Maryland and other places that use DREs, with consequent disruption of the voting process. The Ohio Secretary of State has expressly directed Ohio election workers to use paper ballots to relieve congestion caused by DREs [27], and Indiana and California are similarly prepared. Unfortunately, Maryland is not, as is the case with a number of other states [28].

Table 2. Process time vs. number of voters per voting system (scanner or e-pollbook) that would cause 1% of precincts to have maximum waits over 30 minutes.

| Voter cycle time (seconds) | Actual voters per voting system |
|---|---|
| 5 | 8000 |
| 10 | 3911 |
| 30 | 1074 |
| 60 | 462 |